\documentclass[twocolumn]{IEEEtran}
\usepackage[T1]{fontenc}
\usepackage{ifthen}
\usepackage[utf8]{inputenc}
\usepackage{amssymb}
\usepackage{lmodern}
\usepackage{microtype}
\usepackage{amsmath,amssymb}
\usepackage{arydshln}
\usepackage{graphicx}
\usepackage{dblfloatfix}
\usepackage{booktabs,array,tabularx}
\usepackage{siunitx}
\usepackage{xcolor}
\usepackage{hyperref}
\usepackage[capitalise,noabbrev]{cleveref}
\usepackage{authblk}
\usepackage{caption}
\usepackage{cite}
\usepackage{balance}
\usepackage{multirow}
\usepackage{titlesec}
\usepackage{booktabs}

\titlespacing*{\section}{0pt}{1.5ex plus .2ex minus .2ex}{1.2ex plus .1ex}
\titlespacing*{\subsection}{0pt}{1ex plus .2ex minus .2ex}{0.6ex plus .1ex}

\title{Improving DNS Exfiltration Detection via Transformer Pretraining}
\author[1]{Miloš Tomić}
\author[1]{Aleksa Cvetanović}
\author[1]{Predrag Tadić}

\affil[1]{School of Electrical Engineering, University of Belgrade, Serbia

\{tomicmilos03, aleksa.cvetanovic99\}@gmail.com, tadicp@etf.bg.ac.rs}
\begin{document}
\maketitle

\begin{abstract}
We study whether in-domain pretraining of Bidirectional Encoder Representations from Transformer (BERT)
model improves subdomain-level detection of exfiltration at low false positive rates.
While previous work mostly examines fine-tuned generic Transformers, it does not aim to isolate the effect of pretraining on the downstream task of classification. To address this gap, we develop a controlled pipeline where we freeze operating points on validation and transfer them to the test set, thus enabling clean ablations across different label and pretraining budgets. Our results show significant improvements in the left tail of the Receiver Operating Characteristic (ROC) curve, especially against randomly initialized baseline. Additionally, within pretrained model variants, increasing the number of pretraining steps helps the most when more labeled data are available for fine-tuning.

\end{abstract}

\section{Introduction}

The Domain Name System (DNS) is a common covert channel for data exfiltration as
queries routinely traverse network boundaries and are weakly authenticated or authorized.
Classical detectors mostly rely on hand-crafted features of individual queries (e.g., string length, per-character entropy, number and shape of the labels) or on streaming statistics \cite{ziza, ozery2023informationbasedheavyhittersrealtime}. While such methods are effective at catching high throughput exfiltration, recent studies show that they are vulnerable to low-rate and ``slow'' tunneling, especially in the case of adversaries mimicking benign lexical statistics \cite{dolos}. These findings motivate sequence models that would learn structure directly from subdomains \cite{netgpt,flowtransformer, domurlbert}, without expensive feature-engineering, and raise a central question for this paper: \emph{Does domain-specific masked language modelling (MLM) pretraining of a character-level BERT encoder causally improve detection of DNS exfiltration compared to training randomly initialized models?}

We answer this by building a controlled pipeline that (i) curates and normalizes two large subdomain corpora, (ii) performs \emph{in-domain} character-level MLM pretraining, and (iii) fine-tunes the same encoder for binary subdomain detection. To isolate the effect of pretraining, the main ablations are constructed under the same number of gradient updates against otherwise identical encoders. In order to preserve observed frequencies, training set retains duplicates, whereas validations and test set are string-deduplicated to measure model generalization.
Evaluation uses \emph{frozen operating points}, that is -- thresholds $\tau_\alpha$ are chosen on validation to satisfy $\mathrm{FPR}\!\le\!\alpha$ for $\alpha\!\in\!\{1\%,0.1\%\}$ and then are applied unchanged on the test set. We report normalized partial Area Under the ROC curve across the left tail defined by $\alpha$ ($[0,\alpha]$), Recall at threshold $\alpha$, and calibration (Brier score). Our main ablation isolates the effect of \emph{domain-matched} pretraining by comparing against an otherwise identical randomly initialized encoder. For context only, we include a cross-corpus pretrained model (dataset~B) and briefly verify distributional differences between these two datasets (length/depth/entropy).

\section{Related Work}
Early production systems rely on human-engineered lexical and header features (length, entropy, label count, digit ratio, etc.) and apply conventional machine learning (ML) techniques, or anomaly detection over data streams. Other methods such as sliding-window detectors are effective at flagging high-volume exfiltration with low false-positive rates, however, their performance degrades for slow tunneling. \cite{ziza}

On the other hand, Generative Adversarial Networks (GANs) have been tasked with synthesis of DNS-like strings that mimic benign statistics while encoding malicious payloads. As such, these models substantially reduced the accuracy of state-of-the-art feature-based detectors, further underscoring the brittleness of human-engineered features \cite{dolos}.

Additionally, Transformers have been adapted to tokenized DNS strings and network-flow sequences. These hybrid models augment attention encoders with 1-D CNN blocks or concatenate numerical, or tabular features to token embeddings, therefore yielding improvements over n-gram and CNN baselines \cite{licnn}.

More broadly, graph-enriched approaches that augment BERT encoders such that they rely on similarity-based classification in the learned embedding space, have been shown to achieve higher accuracy when graph context is available \cite{dombertgraph}. Finally, Transformers trained on packet/flow sequences demonstrate that self-supervision on network data is viable even without manual feature extraction. \cite{netgpt,flowtransformer}

Although recent studies indicate that even modest, in-domain pretraining can produce better representations than large, generic models, they typically (i) treat full URLs as independent texts or (ii) omit comparisons to randomly-initialized training under identical architectures \cite{domurlbert}.

We aim to bridge this gap by quantifying the effect of \emph{subdomain-level} MLM pretraining on exfiltration detection under controlled training and evaluation pipeline.

\section{Methods and Setup}

\subsection{Data Processing and Metrics}
\label{methods}

We use two sources: (A) a 24h Internet Service Provider (ISP) DNS request log from a Serbian national ISP augmented with controlled and synthetic DNS exfiltration traces (e.g., \texttt{iodine}, \texttt{DNSExfiltrator}) \cite{ziza2023dns}; and (B) Duck's Party monthly web-crawl subdomains (unique counts with frequencies) \cite{nyuuzyou_subdomains_2025}. For MLM pretraining we deduplicate at the string level (A: \num{590238} uniques; B: \num{13382660} uniques). On the deduplicated corpora, dataset A has longer and deeper subdomains with higher per-character entropy than B (mean length 33.9 vs.\ 22.2; depth 3.16 vs.\ 2.82; entropy 3.63 vs.\ 3.33 bits). Additionally, two-sample Kolmogorov-Smirnov tests on length/labels/entropy reject equality (D=0.26-0.28, $p\!\ll\!10^{-3}$), and we also highlight low lexical overlap: 2.64\% of unique strings are shared.

\textbf{Normalization.} For MLM, we extract subdomains (\texttt{tldextract}), lowercase, strip invalid 
entries, and deduplicate. Splits are partitioned into 80\%/10\%/10\% (train/validation/test) fractions. 

\textbf{Classification set (derived from dataset A).}
We retain rows with valid binary labels $y \in \{0, 1\}$. Additionally, we \emph{retain duplicates in training} to preserve the empirical per-query distribution, which represents the distribution a deployed detector would see. In contrast, validation and test splits are \emph{deduplicated at the string level} (each unique subdomain appears at most once) so that the reported metrics reflect generalization to \emph{distinct} subdomains rather than being dominated by repeated copies of the same string. This prevents optimistic bias (e.g., inflated metrics on overweighed duplicates)  and yields split-to-split comparability when the duplicate histogram is heavy-tailed.

\textbf{Train split.}
We construct duplicate groups by exact string matching. Number of raw rows in the train split: $12270029$ are grouped into $455046$ unique string groups. Inflation ratio is $\rho=\frac{12270029}{455046}=26.96$; duplicate share is $96.29\%$. Group sizes are heavily-tailed: p50/p90/p99/max = 1/9/334/10000. Label counts with duplicates retained: benign $\num{12233579}$ (\num{99.703}\%), malicious $\num{36450}$ (\num{0.297}\%). These prevalences are instance-weighted due to duplicates and are not directly comparable to the \emph{deduplicated} validation/test statistics.

\textbf{Validation split (deduplicated).}
After string-level deduplication we have $N_{\mathrm{val}}=\num{56880}$ unique subdomains: benign $\num{54405}$ (\num{95.65}\%), malicious $\num{2475}$ (\num{4.35}\%).

\textbf{Test split (deduplicated).}
$N_{\mathrm{test}}=\num{56880}$ unique subdomains: benign $\num{53615}$ (\num{94.26}\%), malicious $\num{3265}$ (\num{5.74}\%). We construct the test set to be \emph{disjoint} from pretraining train split, to prevent data leakage.

\textbf{Metrics and evaluation.}
For a decision boundary  $\tau\in[0,1]$, the confusion counts $\mathrm{TP}(\tau),\mathrm{FP}(\tau),\mathrm{TN}(\tau),\mathrm{FN}(\tau)$ are standardly defined at operating point $\tau$, where the positive estimates are given by $s_i \geq \tau$, where $s_i$ is the model score (estimated $\Pr(y{=}1\!\mid\!x ; \theta)$). During training we use $\tau = 0.5$.

From $\{(y_i,s_i)\}_{i=1}^N$ on a split, we form the ROC curve. Because of extreme class imbalance, we define two low-FPR metrics to summarize the left tail of the ROC curve at evaluation and to compare the models:
\begin{align}
\mathrm{Recall@}\tau_{\alpha} &= \max_{\tau:\,\mathrm{FPR}(\tau)\le \alpha}\ \mathrm{TPR}(\tau),\\
\mathrm{pAUC@}\alpha \text{(norm)} &= \frac{1}{\alpha}\int_{0}^{\alpha}\mathrm{TPR}(u)\,du,
\end{align}
where $\alpha\in\{1\%,0.1\%\}$. In practice we use trapezoidal integration over interpolated empirical ROC curve, inserting (0,0) if absent, and normalizing by $\alpha$ so that $\mathrm{pAUC}\in[0,1]$.

\textbf{Frozen operating points (validation\,$\to$\,test).}
To avoid test-tuning, we \emph{freeze} thresholds on the validation split and \emph{only} apply them on test. For FPR targets $\alpha\in\{1\%,0.1\%\}$ we compute the thresholds on the validation split as:
\begin{equation}
\tau_\alpha = \underset{\tau:\,\mathrm{FPR}(\tau)\le \alpha}{\arg\max}\ \mathrm{TPR}(\tau).
\end{equation}
At this operating point we report Recall@$\tau_\alpha$ (test), the realized FPR@$\tau_\alpha$ (test), and confusion counts. We also report pAUC over $\mathrm{FPR}\in[0,\alpha]$ (pAUC@$\alpha$, norm). Additionally, we assess calibration with Brier score. \footnote{\url{https://wandb.ai/lohsmi-school-of-electrical-engineering/Improving-DNS-Exfiltration-Detection}}

\subsection{Model Architecture and Training Setup}
We use a character-level BERT model. Raw subdomains are tokenized over DNS-valid characters (a--z, digits, hyphen, underscore, etc.). The encoder has 12 layers with multi-head self-attention, pre-norm residual blocks, hidden size 768, 12 heads, with feed-forward size 3072. The classification [CLS] embedding feeds a binary classifier. For MLM, we use the same architecture but with an output head projecting to the tokenizer vocabulary space. We use cross-entropy for both MLM and classification tasks.

\textbf{Pretraining.} We pretrain the model on the self-supervised MLM task where a fraction of the input tokens are masked and the model is trained to predict omitted tokens. The masking procedure is the same as in the original BERT model: 15\% of the tokens are selected for possible replacement, out of which 80\% are replaced with [MASK], 10\% with a random token, and 10\% are left unchanged. We pretrain the model on the in-domain corpus for 37.5k and 75k steps (PT-37.5k, PT-75k). We also pretrain the same architecture (HF-PT-37.5k) on a different dataset (B) to measure the effect of transfer from a larger, more heterogeneous corpus and domain mismatch.

\textbf{Fine-tuning.} Finally, we fine-tune the pretrained models on the binary classification task using the labels from corpus A. To isolate the effect of pretraining, we also train a randomly initialized model under the same conditions. We fine-tune all models for 112.5k steps (FT=112.5k), except the randomly initialized one which we train for 150k steps (FT=150k) to measure the impact under the same number of gradient updates. We experiment with three fractions of the labeled data: 10\%, 25\%, and 50\% (stratified by label) alongside 100\% to measure label efficiency of pretraining (fractions indicate the fraction of the $\approx$ 12.3M labels used).

\textbf{Optimization.} Briefly, we use AdamW optimizer with 1\% weight decay, linear warmup (1\% of total steps), and learning rate \num{5e-5}. Batch size is 64 for both MLM and classification tasks. Hardware: 1$\times$ NVIDIA A100 (40\,GB).

\begin{figure*}[t]
    \centering
    \includegraphics[width=0.9\textwidth]{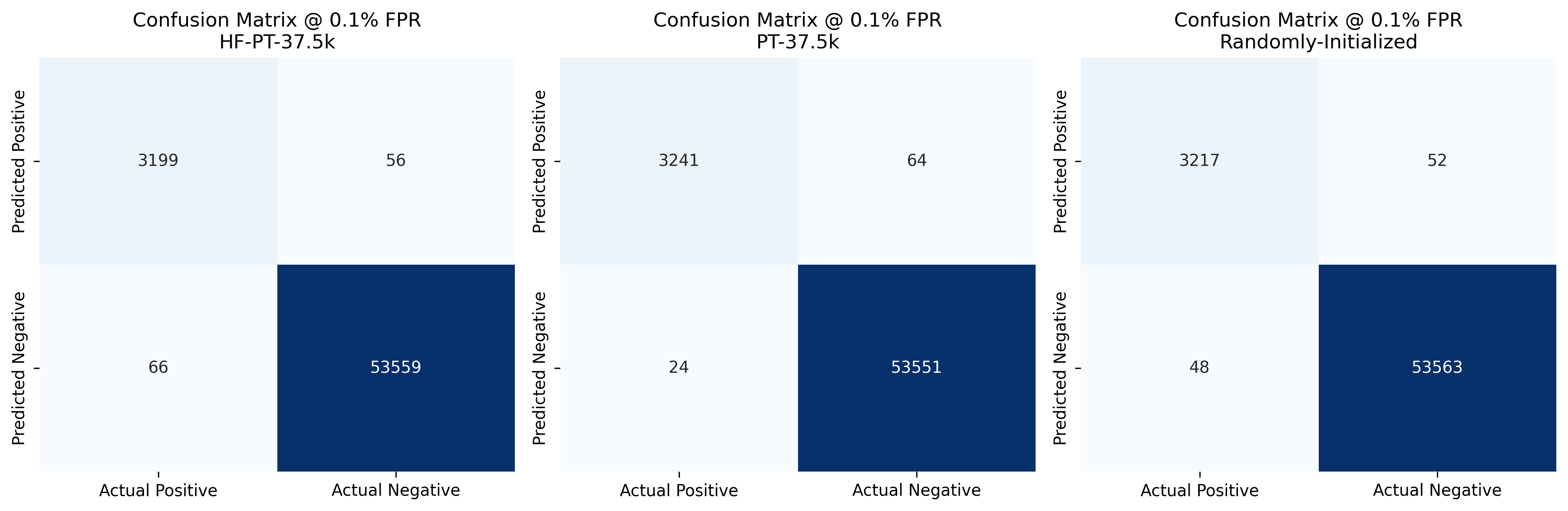}
    \caption{Confusion matrix heatmaps at 0.1\% FPR for Randomly initialized, PT-37.5k and
    HF-PT-37.5k (full-data training).}
    \label{fig:cm_heatmaps}
\end{figure*}

\section{Results}

At the eval-frozen 0.1\% FPR operating point, \emph{in-domain} pretraining yields the highest malicious-class recall and the best calibration, while the cross-corpus model (HF-PT-37.5k) underperforms the randomly initialized baseline, thus highlighting the importance of domain match.
As visualized in Fig.~\ref{fig:cm_heatmaps}, PT-37.5k converts many false negatives into true positives at the cost of a modest increase in false positives at this strict threshold. Low-tail discrimination follows this trend: Table~\ref{tab:lowfpr_test_100} reports higher pAUC@0.1\% and pAUC@1\%, alongside stronger recall at the frozen operating point.
Additionally, PT-37.5k model also yields the best calibration (Brier $9.7\times10^{-4}$ vs.\ $1.3\times10^{-3}$ for Randomly Initialized and $2.2\times10^{-3}$ for HF-PT-37.5k).

\label{sec:100k}

For completeness, we performed an experiment with 100k steps of pretraining on the dataset B (HF-PT-100k), and the results were comparable to the randomly initialized model. Therefore, we further omit the HF model family for brevity, as the experiments suggest that in case of the number of steps constraint, random initialization is preferable. Specifically, the model attains pAUC@0.1\% = 0.9714 and pAUC@1\% = 0.9948.

\begin{table}[h]
    \centering
    \resizebox{\linewidth}{!}{%
        \begin{tabular}{lcccc}
            \toprule
            Model & pAUC@0.1\% (norm) & pAUC@1\% (norm) &
            Recall@$\tau_{0.1\%}$ & Recall@$\tau_{1\%}$ \\
            \midrule
            PT-37.5k & \textbf{0.9830} & \textbf{0.9951} &
            \textbf{0.9926} & \textbf{0.9979} \\
            HF-PT-37.5k & 0.9650 & 0.9896 & 0.9798 & 0.9942 \\
            Randomly Initialized & 0.9790 & 0.9937 & 0.9853 & 0.9966 \\
            \bottomrule
        \end{tabular}%
    }
    \caption{Low-FPR comparison on the held-out test set (100\% train fraction).} 
    \label{tab:lowfpr_test_100}
\end{table}

\begin{table}[t]
    \centering
    \resizebox{\linewidth}{!}{%
        \begin{tabular}{lccccccc}
            \toprule
            \textbf{Frac.} & $\Delta$\,\textbf{pAUC@0.1\%} & $\Delta$\,\textbf{pAUC@1\%} & $\Delta$\,\textbf{FPR@$\tau_{1\%}$} & $\Delta$\,\textbf{TP@$\tau_{1\%}$} & $\Delta$\,\textbf{FP@$\tau_{1\%}$} \\
            \midrule
            10\%  & +0.1004 & +0.0119 & +0.0042 & +13 & +223 \\
            25\%  & +0.0751 & +0.0143 & $-0.0022$ & +14 & $-117$ \\
            50\%  & +0.0418 & +0.0129 & $-0.0036$ & +17 & $-194$ \\
            100\% & +0.0040 & +0.0015 & $-0.0003$ & +4 &  $-18$ \\
            \bottomrule
        \end{tabular}%
    }
    \caption{Label efficiency: PT-37.5k $-$ randomly initialized model on the held-out-test set (thresholds frozen on validation). 
    Positive $\Delta$ means \emph{pretrained} $>$ \emph{random}.}
    \label{tab:label_efficiency_deltas}
\end{table}

\subsection{Label Efficiency}
Varying the supervision budget (10\%, 25\%, 50\%, 100\%), domain-matched pretraining consistently improves low-tail discrimination and probability quality. Compared to a randomly initialized encoder, PT-37.5k raises normalized area under the ROC curve in the $\text{FPR}\!\in[0, \alpha]$ tail consistently over both thresholds $\alpha \in \{0.1\%, 1\%\}$ at 10\%, 25\%, 50\%, and 100\% label fraction levels, respectively (Table~\ref{tab:label_efficiency_deltas}). We also note recall improvements at both operating points indicated by the positive difference in True Positives (TP). Moreover, we observe decrease in Brier score (improvement) in every case ($\Delta \text{Brier}\in[-0.0020; -0.0003]$), indicating better calibration in addition to stronger discrimination.

Threshold transfer (the difference between validation FPR and observed FPR on the test set) is relatively stable across the budget range, except under extreme scarcity (10\% labels), where we observe a small increase in realized FPR ($\Delta\text{FPR@1\%} = {+}0.0042$), trading \mbox{${+}13$ TP} for \mbox{${+}223$ FP} at the same fixed operating point (Table~\ref{tab:label_efficiency_deltas}). Crucially, once the label budget reaches 25\%-50\%, pretraining delivers strictly better performance at the same operating point, both higher recall and \emph{lower} realized FPR with substantial confusion matrix gains (e.g., at 50\%: \mbox{${+}17$ TP} and \mbox{${-}194$ FP}). Benefits persist, albeit smaller, at full-data training (100\% labels). Together with the 0.1\% FPR results in Fig.~\ref{fig:cm_heatmaps}, these findings suggest that domain-matched pretraining gives the largest boost when labels are scarce, while remaining competitive even in the full-data regime.

\subsection{Pretraining Budget}

\begin{table}[h]
    \scriptsize
    \centering
    \resizebox{\linewidth}{!}{%
        \begin{tabular}{l l c c}
        \toprule
                Train frac. &                     Metric &        PT-37.5k &          PT-75k \\
        \midrule
        \multirow{4}{*}{100\%} &          pAUC@0.1\% (norm) &          0.9830 & \textbf{0.9892} \\
                            &            pAUC@1\% (norm) &          0.9951 & \textbf{0.9967} \\
                            & Recall@$\tau_{1\%}$ (test) &          0.9979 & \textbf{0.9982} \\
                            &    FPR@$\tau_{1\%}$ (test) &          0.0086 & \textbf{0.0076} \\
        \midrule
        \multirow{4}{*}{50\%} &          pAUC@0.1\% (norm) &          0.9845 & \textbf{0.9866} \\
                            &            pAUC@1\% (norm) &          0.9960 & \textbf{0.9965} \\
                            & Recall@$\tau_{1\%}$ (test) &          0.9975 & \textbf{0.9985} \\
                            &    FPR@$\tau_{1\%}$ (test) & \textbf{0.0053} &          0.0093 \\
        \midrule
        \multirow{4}{*}{25\%} &          pAUC@0.1\% (norm) &          0.9866 & \textbf{0.9867} \\
                            &            pAUC@1\% (norm) &          0.9959 & \textbf{0.9967} \\
                            & Recall@$\tau_{1\%}$ (test) &          0.9979 & \textbf{0.9991} \\
                            &    FPR@$\tau_{1\%}$ (test) & \textbf{0.0077} &          0.0092 \\
        \midrule
        \multirow{4}{*}{10\%} &          pAUC@0.1\% (norm) & \textbf{0.9650} &          0.9628 \\
                            &            pAUC@1\% (norm) &          0.9902 & \textbf{0.9921} \\
                            & Recall@$\tau_{1\%}$ (test) & \textbf{0.9972} &          0.9963 \\
                            &    FPR@$\tau_{1\%}$ (test) &          0.0112 & \textbf{0.0080} \\
        \bottomrule
        \end{tabular}%
    }
    \caption{Pretraining-budget scaling on the held-out test set.}
    \label{tab:pt_budget_scaling}
\end{table}

At a high level, increasing the pretraining budget from 37.5k steps to 75k steps generally strengthens low-FPR discrimination, and this improvement becomes clearer as the label budget increases. In Table~\ref{tab:pt_budget_scaling}, the 75k-step-pretrained variant tends to outperform the 37.5k-step-pretrained model on both pAUC@0.1\% (extremely-low FPR) and pAUC@1\%, with the gap most visible at 100\% level. This is the regime where larger pretraining budget can be fully leveraged by the downstream supervised task, yielding better metrics in the tail of the ROC curve. 

At 10\%-25\% labels levels, the differences are smaller and occasionally mixed, which reflects the intuition that the label-scarce scenario heavily influences the difference between FPR on validation and test set, as other metrics, and can therefore yield mixed results during the model evaluation.

Operationally, the realized test metrics at the $\tau_{1\%}$ threshold show a nuanced trade-off at low label budgets and clearly improved performance at higher ones. With 50\%-100\% labels, the 75k model typically delivers higher recall at the same operating point, and at 100\% it also reduces realized FPR. At 25\% and 50\%, improvements in recall can come with modest increase in realized FPR. However, pAUC increases at both levels, and as such indicates better average performance in the left tail of the ROC curve.

At 10\% level, we observe the trade-off -- namely, one model is slightly better on recall, while the other is better in terms of the realized FPR, thus indicating the sensitivity of the validation and test FPR difference due to smaller amount of labels. The same pattern holds for pAUC@0.1\% and pAUC@1\%. Taken together, these results suggest that longer pretraining is especially beneficial under higher label budgets, while at very low lable budgets the benefits of longer pretraining are mixed and depend on metric of interest.

\section{Conclusion} 

We showed that character-level BERT encoders \emph{pretrained in-domain} with MLM materially improve DNS-exfiltration detection, especially in the tail of the ROC curve. During our experiments, where we freeze operating points on validation (for $\alpha\!\in\!\{1\%,0.1\%\}$) and transfer them to test, in-domain pretraining delivers higher recall at the same budget, stronger tail separation (normalized pAUC@$\alpha$), and better calibration (Brier) than randomly initialized models. The results show that gains are largest under scarce labels, yet persisting at full data. By 25\%-50\% label budgets the metrics generally favor pretraining, with the 10\%-label caveat, where we record slightly higher realized test FPR traded for more true positives.

As shown in \cref{sec:100k}, cross-corpus pretraining on a different subdomain distribution is comparable to random initialization, but does not match the in-domain gains, thus indicating the value of domain match. Scaling the pretraining budget has been shown to further improve tail metrics when more labels are available.

Overall, domain-matched self-supervision seems to be a label-efficient path to robust DNS exfiltration detection at very low FPRs, attaining superior results when compared to the randomly-initialized baseline.

\section{Acknowledgements}
The authors kindly acknowledge the Government Data Center, Kragujevac, Serbia for providing
access to its HPC cluster.

Predrag Tadić received funding from the taxpayers of the Republic of Serbia, 
through the Ministry of Science, Technological Development and Innovation, under 
contract number 451-03-136/2025-03/200103.

\bibliographystyle{IEEEtran}
\bibliography{refs.bib}

\end{document}